\documentclass[10pt,conference]{IEEEtran}
\IEEEoverridecommandlockouts
\usepackage{cite}
\usepackage{amsmath,amssymb,amsfonts}
\usepackage{algorithmic}
\usepackage{graphicx}
\usepackage{textcomp}
\usepackage[table]{xcolor}
\usepackage[T1]{fontenc}

\usepackage{tikz}
\usetikzlibrary{shapes.geometric, arrows, positioning}

\usepackage{adjustbox}
\usepackage{booktabs}
\usepackage{makecell}

\usepackage{hyperref}

\usepackage{tcolorbox}

\usepackage{enumitem}

\usepackage{lipsum}

\def\BibTeX{{\rm B\kern-.05em{\sc i\kern-.025em b}\kern-.08em
    T\kern-.1667em\lower.7ex\hbox{E}\kern-.125emX}}
\begin{document}

\title{Addressing Quality Challenges in Deep Learning: The Role of MLOps and Domain Knowledge\thanks{This work is partially supported by the GAISSA project TED2021-130923B-I00, funded by MCIN/AEI/10.13039/501100011033 and the European Union Next Generation EU/PRTR. It is also partially funded by the Joan Or{\'o} pre-doctoral support program (BDNS 657443), co-funded by the European Union.}
}

\author{\IEEEauthorblockN{Santiago {del Rey}, Adri{\`a} Medina, Xavier Franch, Silverio Mart{\'i}nez-Fern{\'a}ndez}
\IEEEauthorblockA{
    Universitat Polit{\`e}cnica de Catalunya (BarcelonaTech)\\
    \{santiago.del.rey, xavier.franch, silverio.martinez\}@upc.edu, adria.medina.diez@estudiantat.upc.edu
}
}

\maketitle

\begin{abstract}
Deep learning (DL) systems present unique challenges in software engineering, especially concerning quality attributes like correctness and resource efficiency. While DL models excel in specific tasks, engineering DL systems is still essential. The effort, cost, and potential diminishing returns of continual improvements must be carefully evaluated, as software engineers often face the critical decision of when to stop refining a system relative to its quality attributes. This experience paper explores the role of MLOps practices---such as monitoring and experiment tracking---in creating transparent and reproducible experimentation environments that enable teams to assess and justify the impact of design decisions on quality attributes. Furthermore, we report on experiences addressing the quality challenges by embedding domain knowledge into the design of a DL model and its integration within a larger system. The findings offer actionable insights into the benefits of domain knowledge and MLOps and the strategic consideration of when to limit further optimizations in DL projects to maximize overall system quality and reliability.
\end{abstract}

\begin{IEEEkeywords}
SE4AI, MLOps, software engineering, artificial intelligence, quality attributes
\end{IEEEkeywords}

\section{Introduction}
Deploying deep learning (DL) systems poses significant challenges, particularly regarding correctness, time efficiency, and resource utilization. Achieving these quality attributes (QAs) is crucial for domains like Edge AI, where tasks are executed directly on devices to enable real-time decision-making and minimize dependence on centralized processing. Unlike traditional cloud environments, edge devices operate under constrained resources, requiring optimized models and efficient deployment strategies to meet performance standards without overburdening the system.
 
MLOps (Machine Learning Operations) seeks to mimic DevOps practices and adapt to the specific needs of ML~\cite{lwakatareDevOpsAIChallenges2020}. By supporting tasks like experiment tracking, model versioning, and automatic performance monitoring and reporting, MLOps provides transparency, reproducibility, and streamlined adaptation during the DL life cycle. This is even more relevant with the increasing interest in the fast deployment of DL systems where we must adhere to multiple quality requirements while constant changes are made to models, data, and source code.

Integrating domain knowledge into system design can also significantly improve QAs by guiding the system to make informed decisions with fewer computational demands. Domain-aware algorithms can achieve higher accuracy and efficiency by leveraging context-specific rules or constraints, which complement the data-driven nature of DL models. Nevertheless, such integration remains a challenge in DL systems~\cite{carterAdvancedResearchDirections2023}.

This paper is motivated by the misconception that pre-trained models will work with little effort. In this work, we share our experiences moving away from this partial viewpoint when building DL systems. Specifically, we share key insights from integrating MLOps practices and domain knowledge into a DL system under development, focusing on how these approaches can improve critical QAs. Aligning with \cite{uchitelScopingSoftwareEngineering2024}, our main contribution is a practical reference to improve QAs of a DL system through the use of MLOps and domain knowledge.

\textbf{Data availability statement}: All measurements and analysis are available through the replication package\cite{replication-package}.

\section{Related work}

\textbf{MLOps}. According to John et al.'s MLOps maturity model, there are five stages to reach a successful MLOps implementation~\cite{johnAdvancingMLOpsAd2023}. In this model, experiment tracking and model monitoring are essential to reach the third level of maturity (i.e., manual MLOps). A survey of data science and machine learning practitioners conducted by M{\"a}kinen et al.~\cite{makinenWhoNeedsMLOps2021} shows that tracking and comparing experiments is one of the challenges when implementing MLOps practices. Recupito et al.~\cite{recupitoMultivocalLiteratureReview2022a} conduct a multi-vocal literature review where they analyze the most popular MLOps tools. They identify 13 MLOps tools that can help overcome these challenges. The relevance of MLOps practices is also highlighted in education. For instance, Lanubile et al. propose a project-based approach to teaching MLOps~\cite{lanubileTeachingMLOpsHigher2023a, lanubileTrainingFutureMachine2024}. With this in mind, we report our experiences implementing MLOps so future AI/ML engineers can learn from them.

\textbf{Domain knowledge in DL systems}. Research has shown that domain knowledge can improve system quality~\cite{caiIntegratingDomainKnowledge2023, kuangImpactDomainKnowledge2024}. Most current approaches using domain knowledge in DL systems focus on the DL component by altering the input data or the model~\cite{dashReviewTechniquesInclusion2022}. Other works have explored the integration of DL and rule-based methods~\cite{mendeIntegratingDeepLearning2022}. Similar to the latter, we show how to embed such knowledge directly into the system's logic avoiding overcomplicated DL model design.

\begin{figure*}[!htb]
    \centering
    \includegraphics[width=\linewidth]{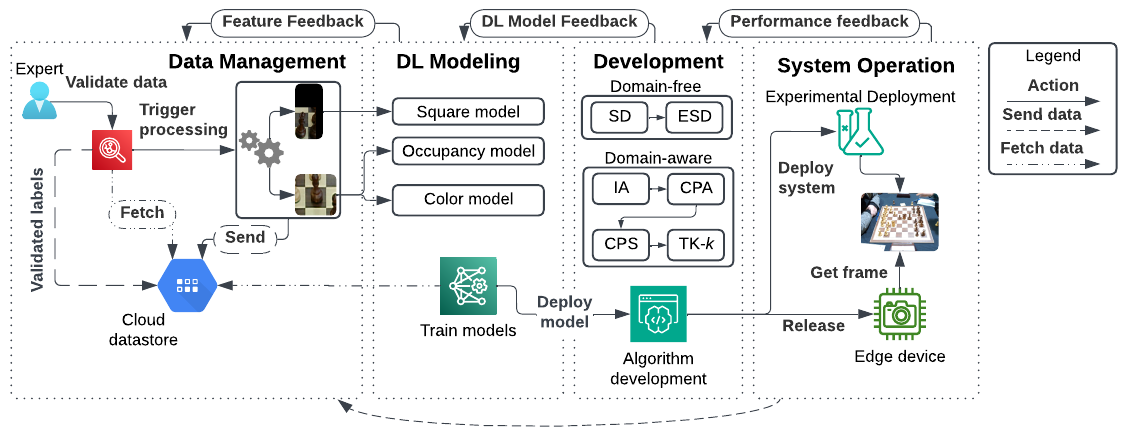}
    \caption{Software development life cycle of the DL system.}
    \label{fig:system-diagram}
\end{figure*}

\section{Study design}
\subsection{Research Goal and questions}\label{sec:RQs}
Following Goal Question Metric (GQM)~\cite{Basili1994TheGQ}, we define our goal as:
\textbf{Analyze} the use of MLOps and domain knowledge \textbf{for the purpose of} improving and monitoring \textbf{with respect to} correctness, time efficiency, and resource utilization \textbf{from the point of view of} AI/ML engineers \textbf{in the context of} a DL system for image recognition.

Then, we derive the following Research Questions (RQs):

\begin{itemize}
    \item {RQ1:} How do MLOps practices impact system quality attributes?
    \item RQ2: To what degree does the use of domain knowledge information improve DL system correctness, time efficiency, and resource utilization?
\end{itemize}

The motivation of RQ1 is to study the benefits and drawbacks of integrating MLOps practices into the life cycle of a DL system and explore the challenges practitioners may face when performing such integration. We will focus on how to perform experiment tracking and energy monitoring during the development phase, and the importance of automatic data collection in operations.

With RQ2 we analyze whether wrapping a DL model within an algorithm encoding domain knowledge improves system QAs. Specifically, we focus on correctness (RQ2.1), time efficiency (RQ2.2), and resource utilization (RQ2.3). We measure these QA using accuracy, prediction latency (hereinafter latency), and energy consumption, respectively. Additionally, we aim to analyze whether algorithms with domain knowledge generally outperform knowledge-free algorithms regarding these QA and in what amount.

\subsection{The DL system for image recognition}

To address our RQs, we choose a DL system developed by the authors, aimed at live broadcasting chess matches using computer vision. Figure~\ref{fig:system-diagram} provides an overview of its life cycle, inspired by the workflow in~\cite{lwakatareDevOpsAIChallenges2020}, divided into four key processes: (i) \emph{Data Management} stores and processes data for training DL models, adapting as needed based on modeling requirements; (ii) \emph{DL Modeling} defines, trains, and evaluates the model architecture, with feedback loops for adjusting data processing; (iii) \emph{Development} integrates the model within the detection algorithm, with feedback for improving the modeling; (iv) \emph{System Operation} deploys the system experimentally and to end users, gathering metrics to improve performance. Data from chessboard interactions then cycles back to the remote server to enhance the Data Management stage. Stages (i - iii) occur in a remote server while stage (iv) happens in edge devices. This structure enables iterative refinements, ensuring a robust, adaptable system.

\textbf{DL system requirements}. To evaluate the degree of functionality and quality of our system, we define the following main requirements~\cite{habibullahNonfunctionalRequirementsMachine2023}:

\textit{1) Functional Requirements (FRs):} \textbf{FR1} - The system must recognize the state of a chessboard from an image. \textbf{FR2} - The system must keep a record of the moves done in a chess match. \textbf{FR3} - The system must monitor the time and energy required to process an image ([related to: RQ1]).

\textit{2) Constraints (Cs):} \textbf{C1} - The system is deployed in edge devices. \textbf{C2} - The system must be continuously running during a whole tournament day.

\textit{3) Quality Requirements (QRs):} \textbf{QR1} - The system must have a minimum accuracy of 90\% ([related to: RQ1, RQ2.1]). \textbf{QR2} - The system must recognize a chessboard image within 2 seconds ([related to: RQ1, RQ2.2]). \textbf{QR3} - The system must minimize energy consumption so that devices can run without a recharge for at least 12 hours ([related to: RQ1, RQ2.3]).

\subsection{Data analysis}
In RQ1 we report our experiences. In RQ2, we conduct the statistical analysis in three stages. First, we employ basic descriptive statistics, summarized in Table~\ref{tab:results}. Second, we use the Shapiro-Wilk test to assess the normality of our data. The test reveals that latency and energy consumption are not normally distributed. Then, we apply a set of statistical tests depending on the RQ. In RQ2.1, we perform a binomial-binomial test using a Z-test to see if the progressive changes introduced in the IA algorithm significantly improved accuracy. We omit SD and ESD given their 0\% accuracy. We also omit the TK-$k$ since it does not add additional domain knowledge to the algorithm and is limited to improving computation efficiency. In RQ2.2 and RQ2.3, we use the Kruskal-Wallis test to see if differences in latency and energy consumption across algorithms are statistically significant. Additionally, we apply Dunn's test as a post-hoc test to determine which pairs of algorithms present differences and similarities in their distributions.

\begin{table}[tb]
    \centering
    \caption{Overview of collected measures. For SD and ESD we show the median square accuracy in parenthesis.}
    \label{tab:results}
    \adjustbox{width=\columnwidth}{%
    {\rowcolors{3}{gray!10}{white}
    \begin{tabular}{@{}llll@{}}
        \toprule
        \thead{Algorithm} & \thead{Accuracy} & \thead{Median\\latency (s)} & \thead{Median energy\\consumption (J)}  \\
        \midrule
        \textbf{Square Detector (SD)} & 0.00\% (71.88\%) & 0.771 & 42.550 \\
        {\bf\makecell{Ensemble Square \\ Detector (ESD)}} & 0.00\% (79.69\%) & 1.271 & 59.120 \\
        \textbf{Initial Algorithm (IA)} & 78.60\% & \textbf{0.246} & \textbf{24.524} \\
        {\bf\makecell{Combined Probabilities \\ Algorithm (CPA)}} & 79.45\% & 0.434 & 31.369 \\
        {\bf\makecell{Combined Probabilities \\ Special (CPS)}} & \textbf{96.85\%} & 0.519 & 34.261 \\
        \textbf{Top-2 (TK-2)} & 96.80\% & 0.326 & 30.532 \\
        \textbf{Top-3 (TK-3)} & \textbf{96.85\%} & 0.353 & 31.927 \\
        \textbf{Top-4 (TK-4)} & \textbf{96.85\%} & 0.379 & 33.469 \\
        \textbf{Top-5 (TK-5)} & \textbf{96.85\%} & 0.402 & 34.761 \\
        \bottomrule
    \end{tabular}}}
\end{table}

\section{Applying traceability and reproducibility with MLOps (RQ1)}

\subsection{Experiment tracking from new data collected in operation}
When deploying the system in operation, we detected a considerable drop in system accuracy. For this reason, we started collecting data in operation and implemented different feedback loops to help drive the decision-making process at each stage, as depicted in Figure~\ref{fig:system-diagram}. Here we focus on the experimental deployment and the DL modeling feedback. We employed MLflow~\cite{MLflow}, an open-source tool conceived to track ML experiments, to implement these two feedback channels. We used MLflow to track the changes made in the DL modeling stage and how they affected our quality metrics. MLflow makes tracking experiments seamless since it has built-in functions that automatically log the most common parameters and metrics of popular ML libraries (e.g., training loss or the optimizer configuration). In addition to common ML metrics, we also needed to log energy consumption during training and deployment, and latency in the deployment. MLflow simplifies this process since it is similar to a typical software logger, allowing users to log any metric by calling a log function and specifying a key-value pair. For instance, one can aggregate the energy consumption reported by energy profilers and log it to MLflow as the total energy consumption of the experiment. MLflow also allows adding artifacts to experiment runs, which we used to save relevant artifacts, i.e., the trained model, and energy trace files generated by the energy profilers. Experiment tracking can happen on different devices leading to the scattering of the logged data if stored on-device. MLflow solves this by allowing the use of a remote tracking server where all the logged data is sent during the tracking process. After data is collected, it can be further processed in the remote server or directly visualized and compared easily inside the MLflow UI. Specifically, the UI became a useful tool to visualize how our decisions impacted metrics like accuracy or latency. Overall, using MLflow to keep track of the main changes we introduced in the system offered invaluable insights into the decision-making process.

\subsection{Energy monitoring}
A requirement for the system is that it must be energy efficient (cf. {\bf QR3}) since it will be deployed in edge devices (cf. {\bf C1}). To monitor energy, we use the Codecarbon Python library~\cite{codecarbon} since it provides energy measurements for specific code fragments instead of giving measurements at the process level. Its use is very straightforward and provides several options to define the scope of the measurements. In our case, we use it as a context manager for the function of performing chessboard detection. Codecarbon relies on the Intel RAPL interface to obtain CPU energy metrics. We could not get reliable measurements with Codecarbon since we used an AMD CPU incompatible with RAPL. Hence, we used Codecarbon with the AMD {\textmu}Prof tool~\cite{AMDMProf} that can profile the energy consumption of AMD CPUs. When using this tool, it should be considered that the measurements are at the machine level. In {\textmu}Prof, measurements for specific parts of the code can be approximated by logging the start and stop timestamps and extracting the energy measurements of that time frame from the report generated by {\textmu}Prof.

\subsection{Automatic data collection and processing}
At the initial stages of the development, we manually generated, validated, and processed new data for model retraining, which was time-consuming. However, to efficiently implement MLOps we are required to automatize this process. Hence, we have improved the system so that all images collected during system operation are sent to a remote centralized file store. The system also registers the associated chessboard status in a remote database. Then, we use a semi-automatic approach to generate the final images to be fed into our models. First, an expert verifies the detected chessboard status corresponds to its associated image employing an ad-hoc validation application we developed. After the necessary corrections, an automatic labeling and processing job is triggered to generate the final data used by the models. An example of the raw and processed data is shown in Figure~\ref{fig:system-diagram}. The validation step also collects the number of correct and incorrect detections. This data is used to compute performance metrics like accuracy for monitoring purposes, e.g., we trigger an alert in case \textbf{QR1} is not met.

\section{Incorporating Domain Knowledge (RQ2)}

\subsection{From domain-free to domain-aware algorithms}
\textbf{Domain-free algorithms} are the most naive approach, using DL models to solve the recognition task without any information except for the input image. We developed two approaches~\cite{delreyAnalysisModelingTraining2023,medinaBruteForceVs2024}. The first is the Square Detection ({\bf SD}). This method uses a multi-class Convolutional Neural Network (CNN) (Square Model in Figure~\ref{fig:system-diagram}) to classify the 64 chessboard squares into empty or a piece type (e.g., white pawn). The second approach, Ensemble Square Detection ({\bf ESD}), uses a combination of models. First, a binary CNN classifies squares as empty or occupied (Occupancy Model in Figure~\ref{fig:system-diagram}). For all occupied squares, a second binary CNN identifies the color of the piece occupying the square (Color Model in Figure~\ref{fig:system-diagram}). Finally, the square model detects the piece type.

The creation of the \textbf{domain-aware algorithm} followed an iterative process, in which we refined the algorithm implementation in several iterations as is depicted in Figure~\ref{fig:system-diagram}.

The process started with the Initial Algorithm ({\bf IA}) where we encoded some domain knowledge and used contextual information. Specifically, the algorithm uses the occupancy model, knowledge of the previous chessboard state, and the legal moves at that specific state. The main idea of this algorithm is to reduce the number of analyzed squares to those containing pieces allowed to move in the current turn. Then it finds which square has become empty since the previous state. Then, list possible destination squares based on the legal moves allowed for the type of piece that has moved and check which has the highest probability of being occupied.

Based on the performance feedback we obtained from the experimental deployment, we applied three improvements to the algorithm. In the first iteration, we adapted the algorithm to consider the probabilities of the origin and destination squares instead of interpreting them separately. Their combination allows the algorithm to obtain an overall likelihood of a move. We refer to this version as the Combined Probabilities Algorithm ({\bf CPA}). In the second iteration, we implemented a special treatment for captures and castling over CPA, which we abbreviate as {\bf CPS}. Specifically, we introduced the Color model and knowledge of players' turns to detect which piece has been captured in case there is more than one option. In the case of castling, the algorithm substitutes the regular formulation of the combined probabilities with a special one considering the casuistic of this type of move. In the last iteration, we aimed to reduce the number of computations by first selecting the $k$ squares most likely to be empty and then proceeding in the same way as the CPS version. We refer to these algorithms as {\bf TK-\emph{k}}, where \emph{k} is the number of squares.

\subsection{Experimental setting and execution}
Experiments are conducted in an \textit{HP Workstation ZBook Power G10} with an \textit{AMD Ryzen™ 7 PRO 7840HS} CPU, a \textit{NVIDIA RTX™ A1000 Laptop GPU}, and 32GB of RAM.

To minimize the presence of background tasks, we close all non-essential processes and use the terminal to launch the runs. To ensure all runs are under the same conditions of system temperature and CPU/GPU strain, we start with a warm-up task before taking measurements and conduct the executions in batches. Each batch lasts about 15 minutes, with a 4-minute cool-down between batches. To collect the metrics, we use the system to recognize 2,000 distinct samples utilizing each algorithm. We annotate each sample and algorithm's actual and predicted chessboard state, latency, and energy consumption. We use the chessboard states to manually compute the system's accuracy as the fraction of correctly predicted chessboard states and the fraction of squares per chessboard that are correctly classified (``square accuracy'').

Before the statistical analysis, we check the correctness of the collected measures. Here, we identified the presence of outliers. We decided to keep them since further inspection did not suggest they were caused by atypical behavior.

\subsection{The Impact of Domain Knowledge}

\subsubsection{Accuracy (RQ2.1)}
Domain-free approaches, i.e., SD and ESD, perform well at the square level. SD achieves a median of 71.88\% square accuracy, and ESD reports a 79.69\%. However, correctly recognizing a chessboard state with these approaches requires 100\% square accuracy. This causes their 0\% accuracy. Instead, domain-aware approaches obtain remarkable results achieving up to 96.85\% accuracy in some versions. Regarding the degree of improvement in the domain-aware algorithms, the Z-test results suggest that the changes introduced in CPA did not significantly improve IA, with a Z=-0.66 and $p>.25$. Comparing CPA to its improved version CPS, we find a substantial impact in accuracy, with a Z=-17.02 and $p<.001$. This indicates that introducing special treatments into the algorithm boosted the system's accuracy.

\subsubsection{Latency (RQ2.2)}
All approaches have similar latencies except ESD which is over one second. Nevertheless, domain-aware methods consistently outperform domain-free methods. IA is the fastest algorithm with a median latency of 0.264s. Indeed, our analysis showed statistical differences in the latency of the algorithms, with the Kruskall-Wallis test reporting H=11871.68 and $p=0$. The reported eta squared ($\eta^2$) is 0.659, indicating a considerable effect size, suggesting that algorithm implementation significantly impacts latency. Further inspection with Dunn's test reveals statistically significant differences between all algorithms.

\subsubsection{Energy consumption (RQ2.3)}
Domain-free approaches report higher median energy consumption. SD reports 42.55J and ESD 59.12J. There is a noticeable reduction in energy consumption when looking at the domain-aware approaches. Specifically, IA stands out requiring only 24.524J. The results from the Kruskall-Wallis test reveal significant differences in the energy consumption of the algorithms, with H=12058.8 and $p=0$. The large effect size ($\eta^2=0.670$) indicates a substantial impact of algorithm design on energy consumption. Dunn's test reveals some similarities among algorithms. Specifically, it does not find statistical differences in the energy consumption of (CPA, TK-2) and (CPS, TK-4).

\section{Discussion}
\subsection{MLOps and software engineering for DL in practice}
RQ1 focuses on how MLOps practices can be implemented in the life cycle of a DL system to improve transparency and reproducibility. This experience paper shows how tools like MLflow can increase the \textbf{traceability} of experiments conducted in the development stage. These tools can be used easily to map certain design decisions to specific QAs of the system. For instance, we can map the selection of an algorithm implemented in our project to QAs like accuracy (cf. {\bf QR1}), latency (cf. {\bf QR2}), and energy consumption (cf. {\bf QR3}). This has allowed us to compare our different approaches in a clean and structured way, without the need for complex additions to our code or managing several files or databases. Although MLflow makes it easy to track experiments, it is highly oriented to training, requiring more coding in other stages like operation.
Besides traceability, we have also focused on monitoring energy consumption to improve \textbf{QR3}. Monitoring energy consumption in software systems is not an easy task. Software-based tools usually come in handy in this context, as they tend to be more easy to set up compared to hardware-based tools. In this regard, we find that there are several alternatives available with similar functionalities~\cite{Sallou_EnergiBridge_Empowering_Software_2023,noureddine-ie-2022, scaphandre, codecarbon}. In our case, Codecarbon became very useful since it is a Python library that can be integrated into the code with little effort. Codecarbon lets you measure the energy consumption of specific code segments, unlike other alternatives that report measurements at the process level. This degree of granularity is essential to detect inefficiencies in the software. However, like most energy profilers, it relies on the RAPL interface. This makes it difficult to work with some AMD CPUs. In such cases, we recommend complementing it with the AMD {\textmu}Prof tool. A common concern when using these tools is the overhead placed into the system. However, previous studies have found that most software-based tools do not place significant overhead in the system~\cite{jayExperimentalComparisonSoftwarebased2023}.
Monitoring during operations is crucial to keep track of the system's performance. However, this stage can also be exploited to gather new data to feed into the DL model automatically, eliminating the need for manual labeling. However, proper checks should be in place to verify the data quality before processing. We had to implement manual checks since the labels required validation before the final processing. For this, we designed a simple tool that presents the data in a visual format. This tool made the validation process faster and less error-prone than manually inspecting the files. In our experience, manual data-related tasks are adequate for the infant stages of a DL system. Indeed, DL systems require strong and reliable data acquisition and processing pipelines that can abstract data scientists from the burden of manually performing such time-consuming tasks.

\subsection{Improving the system with domain knowledge}
RQ2 investigates how using domain knowledge improves the system's QAs. We started developing our system with the naive thought that we could grab a model already trained on chess data and use it out of the box (cf. {\bf SD}). To our surprise, the results---a 0\% accuracy--- were disastrous. At that point, we thought slightly changing the model and retraining it with new data would suffice. Our results in the validation dataset seemed promising with $>90\%$ of square accuracy. Nevertheless, when the model was integrated into the system and deployed we faced the same results as our first attempt. Indeed, we tested multiple model architectures all giving us the same results. At that point, we faced two options. One was to collect more data and continue refining models that were reporting exceptional accuracies during training and validation. Another was to utilize the model intelligently. From our experience, the first option required a huge effort and we believed we could embed domain knowledge into an algorithm that used a DL model as a central component (cf. {\bf IA}). This was a breakthrough in the development of the system. We moved from 0\% accuracy to 78. 6\% while also reducing latency and energy consumption. This is expected since this new approach considerably reduced the number of predictions the system had to perform. Hence, it reduced the chance of mistakes and the number of computations. After the promising results, we continued refining the algorithm with domain knowledge. This led us to a second breakthrough with the CPS algorithm, where we raised the accuracy above the 95\% and still complied with all our requirements. Seeing the good performance of these algorithms, we tried to go back to our initial approach and combine the new models we created for the domain-aware algorithms (cf. {\bf ESD}). Nevertheless, accuracy remained at 0\% so we discarded the domain-free approach.

Overall, our findings underscore the need for software engineers to make informed decisions and know when to stop working to improve quality metrics. This consideration is particularly relevant in DL processes, where the tendency to maximize a single quality metric can overlook the associated costs. Once a system fulfills all required specifications, it is crucial to evaluate whether further improvements in quality metrics will yield meaningful benefits or merely escalate development time and costs.

\section{Conclusions}
This work shows how to integrate MLOps practices and domain knowledge when developing DL systems. Implementing MLOps provided transparent, reproducible experimentation, enabling efficient mapping of design decisions to quality outcomes. Practitioners can use our experiences to set up experiment tracking and energy consumption monitoring as part of their transition to MLOps. Furthermore, we show that embedding domain-specific insights achieved substantial gains in accuracy, latency, and energy consumption, far outperforming domain-free methods. This approach enhances development efficiency by reducing the need for extensive data and model complexity~\cite{delreyAnalysisModelingTraining2023, vilaModelOptimizationChess2023, mollonMachineLearningPipeline2023, moureAutomaticDetectionData2024, medinaBruteForceVs2024}. In future work, we plan to continue developing the system to make it scalable and more robust. This will be evaluated in production in chess tournaments. We also plan to improve our current MLOps process to adhere to the model in~\cite{johnAdvancingMLOpsAd2023}.

\bibliographystyle{IEEEtran}
\bibliography{IEEEabrv,references.bib}

\end{document}